\documentclass[11pt,twoside]{article}
\usepackage{asp2010}

\resetcounters

\begin{document}

\title{Using Limb-Darkening to Measure the Masses of Red Giants}
\author{Hilding~R.~Neilson$^1$, and John~B.~Lester$^2$
\affil{$^1$Argelander Institute for Astronomy, University of Bonn, Bonn, NRW, Germany}
\affil{$^2$University of Toronto Mississauga, Mississauga, ON, Canada}}

\begin{abstract}
We present a novel method for measuring the masses of evolved stars from their limb-darkening observations parameterized as a linear-plus-square-root function with two coefficients.   The coefficients of the law are related to integrated moments of the intensity and carry information about the extension of the stellar atmosphere, which is correlated to the ratio of the stellar radius to mass,~$R/M$.  Here, we show why the limb-darkening law is related to the $R/M$, and apply this result to limb-darkening observations of the microlensing event EROS-BLG-2000-5. 
\end{abstract}

\section{Introduction}
Limb-darkening is an important tool for the study of stellar astrophysics and is measured using  techniques such as optical interferometry and microlensing.  In this work, we describe how a simple two-parameter limb-darkening (LD) law representing stellar intensities can constrain the fundamental properties of red giants including the stellar mass when combined with these observational techniques.  We consider the LD law
$\frac{I}{2\mathcal{H}} = 1 - A - B + \frac{3}{2}A \mu + \frac{5}{4}\sqrt{\mu},$
where $\mathcal{H}$ is the Eddington Flux, and $A$ and $B$ are free parameters. This law has two properties discussed by \cite{Fields2003}: 1)  the LD law conserves flux by definition, and 2) there is a fixed point $\mu_1$ where {\it{all}} relations computed from model stellar atmospheres intersect.  Here, we exploit the second property to constrain stellar masses from limb-darkening observations.

\section{Fixed Point Source}
\begin{figure}
\begin{center}
\includegraphics[width=7.4cm]{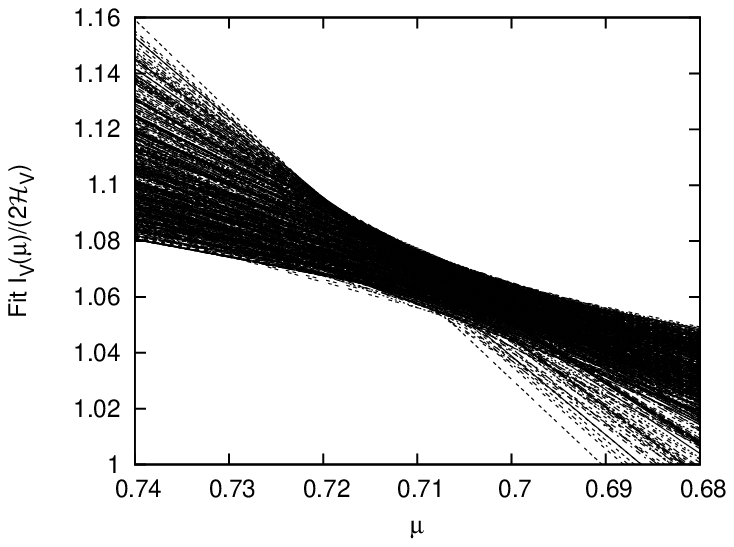}\includegraphics[width=5.9cm]{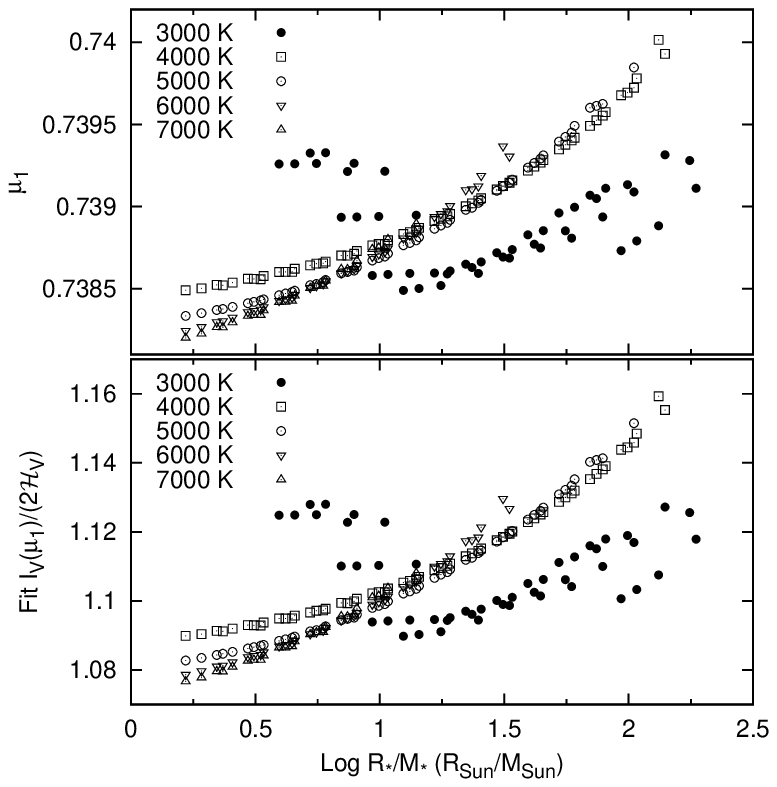}
\end{center}
\caption{(Left) The best-fit V-band LD laws computed for the SAtlas grid of models highlighting the behavior of the fixed point for spherical model atmospheres. (Right) The fixed point, $\mu_1$ and $I(\mu_1)$ as a function of $R/M$ for various $T_{\rm{eff}}$.\label{neilson_f1}}
\end{figure}
The fixed point occurs if $A = \alpha B + \beta$, where $\alpha, \beta$ are constant.  When one fits the two-parameter LD law to actual intensity profiles, the coefficients A and B are linear combinations of the mean intensity, $J$, and $\mathcal{P} \equiv \int I(\mu)\sqrt{\mu}d\mu$.  
For plane-parallel model atmospheres $J = \eta\mathcal{P}$, where $\eta$ is a constant, which is a result of the Eddington approximation.  In spherically symmetric model atmospheres, this behavior changes.

To explore the behavior of the LD law's fixed point, we computed a grid of spherical model stellar atmospheres using the SAtlas code \citep{Lester2008} spanning a parameter range $T_{\rm{eff}} = 3000 - 8000~$K, $\log g = -1 - 3$, and $M = 2.5 - 10~M_\odot$. From Figure~\ref{neilson_f1}, it is clear that the fixed point varies.  This variation is because in spherical symmetry the Eddington approximation is not constant.  Furthermore, we find that the location of the fixed point and the intensity of the LD law at the fixed point is a function of the ratio of the stellar radius and mass, $R/M$, as shown in Figure~\ref{neilson_f1}. 
This result means that we can use the coefficients of the LD law as a measure of $R/M$ for a star.  Thus, knowing the $T_{\rm{eff}}$, $R$ and limb-darkening for a star, we can measure the mass. 

\section{Application to EROS-BLG-2000-5}
Fields et al. (2003) presented LD fits to a microlensing event and compare to stellar atmosphere model in V,I,H-bands and two other bands that we do not consider here.  The LD coefficients can be found in that article.  The lensed star is a K-giant with an estimated distance of $9\pm3~$kpc, angular diameter of $6.62\pm0.58$ $\mu$as, and $T_{\rm{eff}} = 4200~$K \citep{An2002}, and hence $R = 12.8\pm5.4~R_{\odot}$.  Combining this information with the relations shown in Figure \ref{neilson_f1} and the \cite{Fields2003} LD coefficients, we compute the $R/M$  for the V,I, and H-band data to be $20.33^{+2.85}_{-2.59}, 35.21^{+3.25}_{-3.04}, 27.16^{+2.80}_{-2.60}~R_\odot/M_\odot$ and mass,  $M=0.63^{+0.45}_{-0.31}, 0.36^{+0.24}_{-0.17},0.47^{+0.31}_{-0.17}~M_\odot$, respectively.
We note that the values of $R/M$ for each waveband are not consistent.  \cite{Fields2003} noted that the multisite observations of the microlensing event leads to LD fits that vary depending on sampling of the data. Thus the inconsistency found here is related to the probable inconsistency within the observations themselves.

\bibliographystyle{asp2010}

\bibliography{neilson_cite}


\end{document}